# Modeling observations of solar coronal mass ejections with heliospheric imagers verified with the Heliophysics System Observatory

**C. Möstl[1,2]\*, A. Isavnin[3], P. D. Boakes[1,2], E. K. J. Kilpua[3], J. A. Davies[4], R. A. Harrison[4], D. Barnes[4,5], V. Krupar[6], J. P. Eastwood[7], S. W. Good[7], R. J. Forsyth[7], V. Bothmer[8], M. A. Reiss[2], T. Amerstorfer[1], R. M. Winslow[9], B. J. Anderson[10], L. C. Philpott[11], L. Rodriguez[12], A. P. Rouillard[13,14], P. Gallagher[15], T. Nieves-Chinchilla[16] and T. L. Zhang[1]**

[1]Space Research Institute, Austrian Academy of Sciences, Graz, Austria.
[2]IGAM-Kanzelhöhe Observatory, Institute of Physics, University of Graz, Austria.
[3]Department of Physics, University of Helsinki, Helsinki, Finland.
[4]RAL Space, Rutherford Appleton Laboratory, Harwell, Oxford, UK.
[5]University College London, UK.
[6]Institute of Atmospheric Physics CAS, Prague, Czech Republic.
[7]The Blackett Laboratory, Imperial College London, London, UK.
[8]Institute for Astrophysics, University of Göttingen, Göttingen, Germany.
[9]Institute for the Study of Earth, Oceans, and Space, University of New Hampshire, Durham, NH, USA
[10]The Johns Hopkins University Applied Physics Laboratory, Laurel, MD, USA.
[11]Department of Earth, Ocean and Atmospheric Sciences, University of British Columbia, Vancouver, British Columbia, Canada.
[12]Solar–Terrestrial Center of Excellence – SIDC, Royal Observatory of Belgium, Brussels, Belgium.
[13]Institut de Recherche en Astrophysique et Planétologie, Université de Toulouse (UPS), France.
[14]Centre National de la Recherche Scientifique, Toulouse, France.
[15]School of Physics, Trinity College Dublin, Ireland.
[16]Heliophysics Science Division, GSFC/NASA, Greenbelt, MD, USA

Corresponding author: Christian Möstl (christian.moestl@oeaw.ac.at)

**Key Points:**

- Hindsight predictions of CMEs by modeling observations from the STEREO heliospheric imagers science data are analyzed for 2007-2014.

- About 1 out of 4 predicted CME arrivals is observed in situ as a magnetic obstacle by MESSENGER, Venus Express, Wind and STEREO-A/B.

- Although using strong assumptions on CME physics, prediction accuracies for CME arrival times are similar to numerical or analytical models.

## Abstract

We present an advance towards accurately predicting the arrivals of coronal mass ejections (CMEs) at the terrestrial planets, including Earth. For the first time, we are able to assess a CME prediction model using data over 2/3 of a solar cycle of observations with the Heliophysics System Observatory. We validate modeling results of 1337 CMEs observed with the Solar Terrestrial Relations Observatory (STEREO) heliospheric imagers (HI) (science data) from 8 years of observations by 5 in situ observing spacecraft. We use the self-similar expansion model for CME fronts assuming 60 degree longitudinal width, constant speed and constant propagation direction. With these assumptions we find that 23%-35% of all CMEs that were predicted to hit a certain spacecraft lead to clear in situ signatures, so that for 1 correct prediction, 2 to 3 false alarms would have been issued. In addition, we find that the prediction accuracy does not degrade with the HI longitudinal separation from Earth. Predicted arrival times are on average within 2.6 ± 16.6 hours difference of the in situ arrival time, similar to analytical and numerical modeling, and a true skill statistic of 0.21. We also discuss various factors that may improve the accuracy of space weather forecasting using wide-angle heliospheric imager observations. These results form a first order approximated baseline of the prediction accuracy that is possible with HI and other methods used for data by an operational space weather mission at the Sun-Earth L5 point.

## 1 Introduction

A major goal in space weather research is to improve the accuracy of forecasts concerning coronal mass ejection (CME) arrival times and speeds, and whether a CME impacts or misses the Earth. This is needed to ensure that potentially disrupting effects for society are mitigated [*Oughton et al.,* 2017, and references therein]. The heliospheric imagers [HI, *Howard et al.,* 2008; *Eyles et al.,* 2009] onboard STEREO have pioneered this approach for the first time away from the Sun-Earth line. Because solar wind structures such as CMEs can be observed as they propagate towards the Earth and even as they sweep over it [*Davis et al.* 2009*; Möstl et al.* 2010*; Liu et al.* 2010*; Colaninno et al.* 2013*; Möstl et al.* 2014*; DeForest et al.,* 2016*; Wood et al.* 2017], it makes sense to use modeling approaches specifically designed for HI [*Liu et al.,* 2010*; Lugaz et al.,* 2010*; Davies et al.,* 2012*; Möstl and Davies,* 2013*; Möstl et al.,* 2014*; Rollett et al.,* 2016] to predict CME arrival times and whether a CME will impact a certain heliospheric location, similar to analytical or numerical modeling [e.g. *Vršnak et al.,* 2014*; Mays et al.,* 2015]. For assessing the prediction accuracy, data not only of the near Earth solar wind, but also from several other spacecraft operating in the heliosphere can be used. This collection of spacecraft is known as the Heliophysics System Observatory (HSO). Here, we use data from the HSO spacecraft operating in the heliosphere and at the Sun-Earth L1 point, but no data from other spacecraft in geospace.

In this study, we test the validity of using HI observations for space weather forecasting, in particular for the CME arrival time, and the accuracy of hit and miss predictions. These are essential results for a possible future space weather mission to the L5 point, which could continuously monitor the Sun and the space between the Sun and Earth with HI [*Lavraud et al.,* 2016; *DeForest et al.,* 2016]. Wide field imagers, somewhat akin to the STEREO/HI instrument, will also be onboard Solar Orbiter and Solar Probe Plus, and STEREO-Ahead/HI continues to observe the solar wind between the Sun and Earth in July 2015, emphasizing that our results will provide a baseline for future missions.

We present here the first work in which predictions based on a large scale sample of 1337 CMEs observed with either HI instrument (HIA on STEREO-A, and HIB on STEREO-B) are tested with 641 interplanetary coronal mass ejection events observed in situ, obtained between April 2007 and September 2014 and ranging in heliocentric distances from 0.308 to 1.086 AU. It has been shown previously [e.g. *Colannino et al.* 2013; *Möstl et al.,* 2014; *Tucker-Hood et al.,* 2015] that STEREO/HI may enhance the warning times and the accuracy of CME predictions, even in near real time [*Tucker-Hood et al.,* 2015]. However, the conclusion of these studies was that there is still room for improvement concerning the HI modeling and that the maximum number of events that could be ratified with a corresponding in situ observation was around 20 [e.g. *Möstl et al.,* 2014; *Tucker-Hood* et al., 2015]. Note that in this present work, all predictions are in fact made in hindsight.

For predicting space weather at Earth or any other planet it is necessary to test the HI modeling techniques also for distances other than 1 AU. Studies so far have focused on in situ data near 1 AU [*Möstl et al.,* 2014; *Rollett et al.,* 2016] or on data taken in the near-Earth solar wind [*Tucker-Hood et al.*, 2015]. Our aim in this work is thus twofold: (1) to substantially increase the number of CME events available for the statistical prediction performance analyses, and (2) to test the forecasts based on HI data for heliocentric distances other than 1 AU. This is of relevance for studies of planetary space weather and even for manned Mars missions.

## 2 Methods and Data

This work is an outcome of the *HELiospheric Cataloguing, Analyses, and Techniques Service* (HELCATS) project, in which 8 academic research institutions from around Europe were involved from May 2014 - April 2017. During the time of this project, we have established, among many other catalogs from heliospheric datasets, several catalogs that are relevant for this study: the catalog of observed CMEs with HI is called HIGeoCAT, the catalog of predicted arrivals ARRCAT, the catalog of the in situ interplanetary data DATACAT, and information on the in situ CME (ICME) observations is gathered in the ICMECAT. It contains data from the solar wind observatories Wind, STEREO-A, STEREO-B, Venus Express (VEX), MESSENGER, and Ulysses. We use all spacecraft for a comparison to HI except Ulysses because there are too few CME events observed by this spacecraft.

## 2.1 HIGeoCAT

The CME catalogue, HIGeoCAT, forms the basis of our investigation. This is underpinned by a catalog of CMEs visually identified in images from the inner (HI1) camera on each STEREO spacecraft (available on the HELCATS website, see sources of data at the end of this article), using science data (not real-time beacon data). For each CME in that catalog categorized as either "good" or "fair", a time-elongation map (commonly called a J-map) was constructed from combined HI1 and HI2 difference observations, along a position angle corresponding to — as near as practicable — the CME apex. Note that in 71% of cases, this is within 30° of the solar equatorial plane, which is situated at a position angle of 90 (HIA) and 270 degree (HIB). This demonstrates the tendency for CMEs to be confined to low equatorial latitudes. The time-elongation profile corresponding to the leading edge of the CME was extracted, by clicking along the appropriate trace in the J-map [*Sheeley et al.,* 1999; *Rouillard et al.,* 2008]; for each CME, this process was repeated multiple times. The resulting tracks were fitted with the self-similar expansion method [*Davies et al.*, 2012] with 30° half width, in short called SSEF30. Thus, for each CME a 60° full width in the solar equatorial plane is assumed. The method assumes the CME front to be of circular shape, and is based on geometrical aspects of objects that propagate at a given angle to the observer. We use only single-spacecraft HI results, similar to what would be available on an L5 mission, and no stereoscopic results are used. The CME parameters in HIGeoCat include among others a constant speed and constant propagation direction, and a launch time *sse_launch,* which is a back-projection to the time of the CME's inception in the solar corona. It is usually very close in time to the first appearance of the CME in the STEREO coronagraph COR2 [*Möstl et al.*, 2014]. The time range of the HI observations and modeling was 1 April 2007 to 27 September 2014.

## 2.2 ARRCAT

**Figure 1a** shows an overview of the positions of the predicted CME impacts in the heliosphere. This is based on the ARRival CATalogue (ARRCAT) that was produced for all events in HIGeoCAT. ARRCAT contains impacts at Earth L1, STEREO-A, STEREO-B, VEX and MESSENGER. These are further called the targets. We have also produced arrival lists for Mars, Saturn, Ulysses, Mars Science Laboratory (MSL), MAVEN, and Rosetta, but they are not used in this work. The position of VEX is assumed equal to the location of Venus, and the location of MESSENGER is equal to Mercury after orbit insertion on 18 March 2011.

Essentially, for each CME it is assessed whether a planet or spacecraft is currently situated within ± 30 degree heliospheric longitude (Heliocentric Earth Equatorial, HEEQ [*Thompson*, 2006]) of the CME propagation direction. If this is the case, the SSEF30 circle is expected to impact a target, and an entry in ARRCAT is produced. The heliospheric positions of those targets are assessed at the time *sse_launch* as the CME leaves the Sun. For creating a predicted arrival time of a CME, we add a time of travel, depending on the heliocentric distance of the target and the CME speed, to *sse_launch* [for details on the method see *Möstl and Davies*, 2013]. The predicted impact time *target_arrival* and the speed are calculated including the circular shaped SSEF30 front, which results in significant delays of flank impacts compared to head-on hits due to the circular geometry [*Möstl and Davies*, 2013].

At VEX and MESSENGER, the spacecraft position at *sse_launch* is actually a few degrees away from the planet position at the time of the hit. For STEREO-A/B and Earth, this problem does not occur because in the HEEQ coordinate system, their positions move very little (STEREO) or not at all (Earth). Therefore, a slight shift occurs between the longitudinal position of the predicted impact and the actual longitude of the targets VEX and MESSENGER at impact time. For VEX, this amounts on average to $1.8 \pm 0.9$ degree longitude, and for MESSENGER its $4.2 \pm 2.6$ degree.

It is not straightforward to calculate how this shift affects the arrival time calculation, because this depends on many factors such as CME speed, heliocentric distance and in particular the position of the impact along the circular front. However, we can make an estimate based on average values. With the method from *Möstl and Davies* [2013] we find that for an average predicted impact speed of 513 km s$^{-1}$ at the distance of Venus, and a difference to the CME apex of 15 degree the difference in arrival time due to a shift of 1.8 degree is about 1.3 hours. For very extreme cases (very slow CME, far flank hits) it can also be on the order of 10 hours, which is caused by the strong curvature of the circular SSEF shape at the front edges. A similar value of 1.7 hours difference in arrival time is found for MESSENGER, again for average values. This means that the arrival time calculation is very often not much affected by the systematic error in assessing the spacecraft position at launch time. Nevertheless, this is a problem inherent to our current method of arrival time calculation, and should be fixed in future updates.

## 2.3 DATACAT

The name DATACAT is given to the collection of all files that include the in situ magnetic field and plasma data, downloaded from various online sources at the respective sites of the missions, and converted to a single coordinate system. The instrument studies of the magnetometers are for Wind *Lepping et al.* [1995], for VEX *Zhang et al.* [2006], for STEREO *Luhmann et al.* [2008], and for MESSENGER *Anderson et al.* [2007]. For VEX and MESSENGER, only magnetic field data are available. For completeness, plasma speeds, temperatures and densities were taken from STEREO [PLASTIC, *Galvin et al.,* 2008] and Wind [SWE, *Ogilvie et al.,* 1995]. Except for a comparison to the plasma speeds at Wind we do not use the plasma data in this study. All data were either available or resampled to a 1 minute time resolution.

The in situ data start on 1 January 2007. MESSENGER was still in cruise phase and has a few long intervals when no data was taken. The magnetic field data from the other missions are almost continuous. The magnetosphere measurements have been removed from MESSENGER observations in Mercury orbit, starting with the orbit insertion at Mercury on 18 March 2011, based on a manually established list of magnetosphere crossings for every MESSENGER orbit [*Winslow et al.,* 2013, updated until the end of the mission]. For each orbit, we have excluded all data that were taken in between the outer boundaries of the bow shock. We have also manually removed any residual spikes (defined by clear instrumental artefacts of singular measurements above 100 nT) in the magnetic field data so the calculations of ICME parameters are not affected. Calibrated level 2 magnetic field and plasma data from STEREO IMPACT and PLASTIC instruments, respectively, were obtained from the Space Physics Center of UCLA. Wind magnetic field and plasma data from MFI and SWE instruments were obtained from Coordinated Data Analysis Web service maintained by NASA. For removing the induced magnetosphere of Venus from the VEX data, a slightly modified formula from *Zhang et al.* [2008] was used. We use a terminator crossing of 3.5 instead of 2.14 and 2.364 [*Zhang et al.,* 2008], which works to exclude almost all movements of the bow shock while still retaining large amounts of solar wind data.

Due to the end of the missions, VEX data terminates on 26 November 2014, and data from MESSENGER on 30 April 2015. STEREO-A moved to solar conjunction starting with 19 August 2014, and contact with STEREO-B was lost on 27 September 2014. Wind data are available still after than 27 September 2014, but the data used in our study end at this time because no continuous in situ nor imaging data from either STEREO spacecraft are available from this date to July 2015, when STEREO-A continued observations.

## 2.4 ICMECAT

We define an interplanetary coronal mass ejection (ICME) as the full interval of disturbed solar wind, including the shock, sheath, magnetic obstacle, and a possible wake region [*Rouillard,* 2011; *Kilpua et al.*, 2013]. We have gathered lists of in situ CME observations from various online catalogs and published sources, in the timeframe January 2007-December 2015, which overlaps fully with the HIGeoCAT. This results in 668 ICME events. For the comparison with HI, from April 2007 to September 2014, 641 events at the 5 target spacecraft are available. The reduction from 668 events to 641 is caused by the time ranges January 2007 to March 2007 and October 2014 to December 2015 not having continuous HI observations and because the few events observed by Ulysses have been omitted for comparison to HI.

**Figure 1b** provides an overview of the positions where those ICMEs have been detected with respect to Earth, demonstrating the 360° heliospheric coverage of ICMEs. It is clear that merging the individual catalogs leads to some problems concerning different definitions of parameters. To work around this, we have taken *only times* from the individual catalogs: from the Wind ICME catalog for L1 ICMEs (by Teresa Nieves-Chinchilla), from the STEREO-A/B ICME list by Lan Jian, and from the VEX ICME list from *Good and Forsyth* [2016]. ICMEs at MESSENGER were taken from *Winslow et al.* [2015] and *Good and Forsyth* [2016]. We have updated all ICME lists except for Wind to include ICMEs until the end of the missions (VEX, MESSENGER) or until solar conjunction was reached (STEREO-A) or contact was lost (STEREO-B).

There are four times included in ICMECAT, for each event. First, there is the *icme_start_time*, which is equal to the shock arrival or the beginning of a significant density enhancement. Then, *mo_start_time* and *mo_end_time* are given which signify the beginning and end of the magnetic obstacle, which is taken as an umbrella term for any significant magnetic structure with higher total magnetic field, and it may include smoothly rotating or complex magnetic field components. If there is no sheath region, i.e. the ICME starts immediately with a magnetic obstacle, the *icme_start_time* is similar to *mo_start_time*. For Wind only, a separate *icme_end_time* is given, for all other spacecraft the *icme_end_time* is the same as the *mo_end_time*. Interacting ICMEs are treated usually as a single ICME in all catalogs, and only if they can be clearly separated they are considered as two individual structures.

ICMEs were identified at STEREO by Lan Jian using a range of plasma and magnetic field signatures, namely, an enhanced total perpendicular pressure (including the contribution of protons, electrons and alpha particles), a declining speed profile, low proton temperature, a relatively enhanced magnetic field strength, and smooth magnetic field rotations [*Jian et al.*, 2006, *Jian et al.*, 2013]. The presence of at least three of these signatures was required for an ICME identification. A minimum time duration for the signatures was not imposed, although all of the cataloged ICMEs had magnetic obstacle durations of at least 2.5 hours.

The magnetic obstacles of ICMEs at Wind were identified as regions of magnetized plasma where the magnetic pressure was significantly greater than the plasma pressure (i.e., where the plasma-*β* was low). Obstacles that showed rotation in at least one field component were defined as flux ropes, and obstacles that showed no rotation were defined as ejecta. Similarly to the STEREO catalog, signature duration was not used as an identification criterion; the shortest magnetic obstacle duration of an ICME observed at Wind was 5.5 hours.

For MESSENGER and VEX, only magnetic field data were available for identification of ICMEs and subsequent derivation of the parameters, thus they need some special attention. ICMEs observed at MESSENGER and VEX that have been listed by *Good and Forsyth* [2016] were all identified by the presence of magnetic flux rope signatures, i.e., relatively smooth rotations of the magnetic field direction coinciding with an elevated magnetic field strength. In general, flux ropes that had a duration of less than 4 hours were not included in their catalog. No shock-driving identification criterion was imposed. MESSENGER observed ICMEs during both its long cruise phase and subsequent orbital phase at Mercury (from 18 March 2011); all ICMEs identified at VEX were observed during the spacecraft's orbital phase at Venus. ICMEs observed by MESSENGER and VEX during the orbital phases were obscured to varying degrees by magnetospheric intervals, although ICME boundaries were clearly visible in most cases. In case of overlaps of the MESSENGER ICME lists by *Winslow et al.* [2015] and *Good and Forsyth* [2016] we have taken the times from *Winslow et al.* [2015] because they include both the ICME start time and the magnetic obstacle times, whereas in the *Good and Forsyth* [2016] list only the magnetic obstacle times were available. We have also added shock times to the list by *Good and Forsyth* [2016] for those events from VEX and MESSENGER where a clear shock preceded the magnetic obstacle.

For all ICMEs, we have derived 30 parameters (e.g. mean and maximum magnetic field, *Bz* field parameters, minimum variance analysis of the magnetic obstacle) directly from the data in a homogeneous way, thus eliminating the need to compile different parameters from the different catalogs which might differ in their definitions from one catalog to another. In summary, in ICMECAT 483 ICMEs were observed close to 1 AU (Wind, STEREO-A, STEREO-B), and 180 events in the inner heliosphere by VEX and MESSENGER, and 5 events by Ulysses. However, for our analysis only the *icme_start_time*, the mean magnetic field in the magnetic obstacle and the sheath speed at Wind is used for a comparison to the HI predictions. The catalog is thus introduced here also as a basis for future studies.

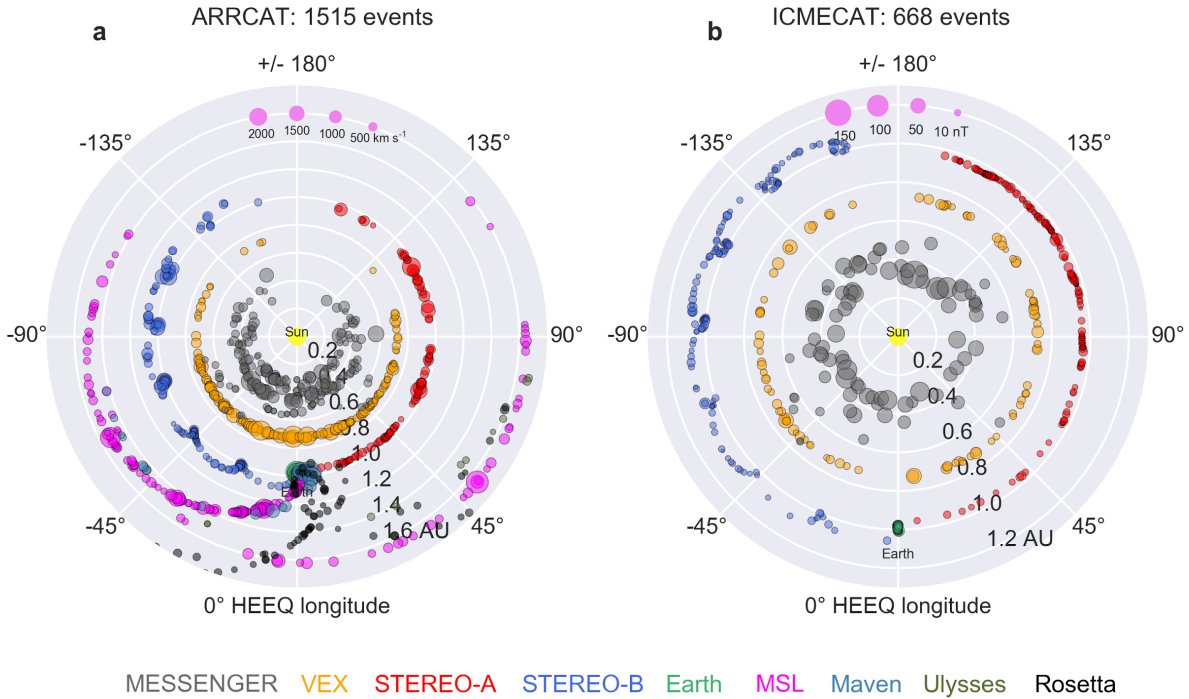

***Figure 1: Overview of ARRCAT and ICMECAT.*** *Both panels show the solar equatorial plane.*
***(a)*** *All CMEs in HIGeoCAT were checked if they potentially arrive at various planets and*
*spacecraft as indicated by the color code at the bottom, based on the shape model of a self-*
*similar expanding circle with 30° half-width (SSEF30). Each dot marks a predicted arrival at*
*MESSENGER, VEX, STEREO-A/B, Earth/L1, MSL, MAVEN and Ulysses and Rosetta. The size*
*of the dot indicates the predicted impact speed, which is an overestimate due to the constant*
*speed assumption of SSEF30.* ***(b)*** *Overview of in situ detections of ICMEs, showing the longitude*
*and radial distance at which the detection happened, as collected in ICMECAT. The size of the*
*circle indicates the mean magnetic field strength in the magnetic obstacle. Here, MESSENGER,*
*VEX, STEREO-A/B and Earth/L1 are shown.*

## 2.5 Summary of data and methods

**Figure 2** is a screenshot of an animation that provides a convenient overview of the spacecraft positions, the CMEs in HIGeoCAT and the magnetic field components and total field for the full duration of the data we study. The movie has a 6 hour time resolution, and covers 2007-2015 (see links to the online animation at the end of the paper). In the left panel, the SSEF30 circles propagate away from the Sun, for the inner heliosphere up to Mars orbit, for all CMEs observed by HI. For each CME, kinematics have been established by calculating $R(t) = V\_sse$ x $t$, where $R(t)$ is the distance to Sun center, $V\_sse$ the speed from SSEF30 and $t$ the time, starting with the SSE launch time. These kinematics are then interpolated to the movie frame times. All apex positions for each CME are then sorted by time. When plotting the movie, for each movie frame, it is checked how many CMEs are currently "active" at that particular moment in time in the spatial domain covering approximately 0 to 2.1 AU. The SSE circle is then plotted for each CME with + /- 110 degrees around the apex position, resulting in a circular arc-like CME front appearance.

Every circle in the movie has a full width of 60°, which was chosen to be consistent with the average CME width in SOHO coronagraph observations [*Yashiro et al.,* 2004]. We know from previous studies [e.g. *Möstl et al.*, 2014] that the accuracy of the CME direction from SSEF30 has some problems - sometimes one can recognize two circles (one red and one blue) which are obviously the same CME seen from STEREO-A and B because of the close timing, but the directions differ sometimes on the order of up to 50 degree in extreme cases although they describe the same CME. In the right panel of the figure we show magnetometer observations by the 5 spacecraft as collected in DATACAT.

In summary, we have established the catalog of CME arrivals that we call ARRCAT, which is based upon the SSEF30 modeling in HIGeoCAT. We have also gathered in situ observations from up to 9 years of 5 different spacecraft in the DATACAT, from which we have derived ICME parameters based on the timings in existing ICME lists from the individual spacecraft, which we have updated and merged together in the ICMECAT.

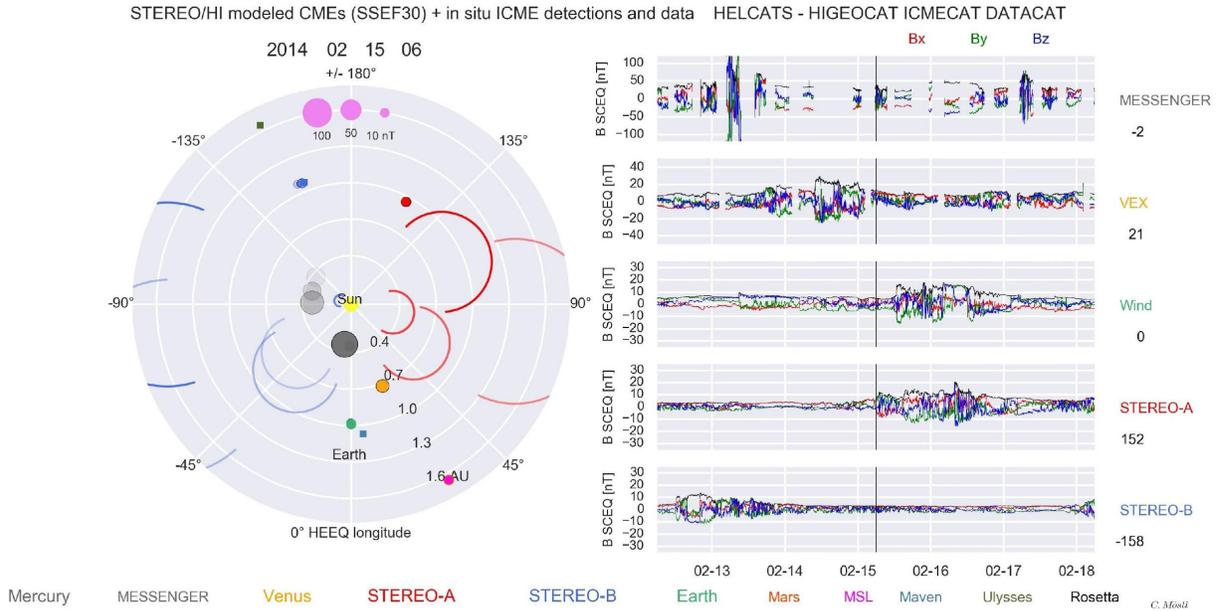

***Figure 2***. ***Screenshot of a visualization of CMEs observed with STEREO/HI and their in situ detection by various spacecraft, covering 2007-2015 at a 6 hour time resolution***. *In the left panel semi-circles denote CMEs that propagate away from the Sun as modeled with SSEF30. Circles that light up and fade around spacecraft positions are actual in situ ICME detections. The size of the circle that lights up is related to the mean magnetic field in the ICME, given by the legend on the upper part. In the right panel, the in situ magnetic field components in colors (Bx red, By green, Bz blue) and the total field (black) are shown from top to bottom for the 5 spacecraft MESSENGER, VEX, Wind, STEREO-A, and STEREO-B. The number below each spacecraft label on the far right is the HEEQ longitude at the current animation time, which is shown above the left panel and also given by the vertical solid line on each plot in the right panel.*

## 3 Results

To test the predictive capabilities of data from STEREO/HI, we have essentially verified the catalog of predicted arrivals (ARRCAT) with respect to the *icme_start_time* in the catalog of in situ ICME detections (ICMECAT). This shows some fundamental results which are important for space weather forecasting with an HI instrument, based on modeling with the SSEF30 technique.

**Hits and false alarms**

In **Figure 3** we assess hit and false alarm predictions: Panel (a) shows the number of predicted CME hits, only from HIA, for each year as shaded bars. Panel (b) shows the same based on HIB observations.

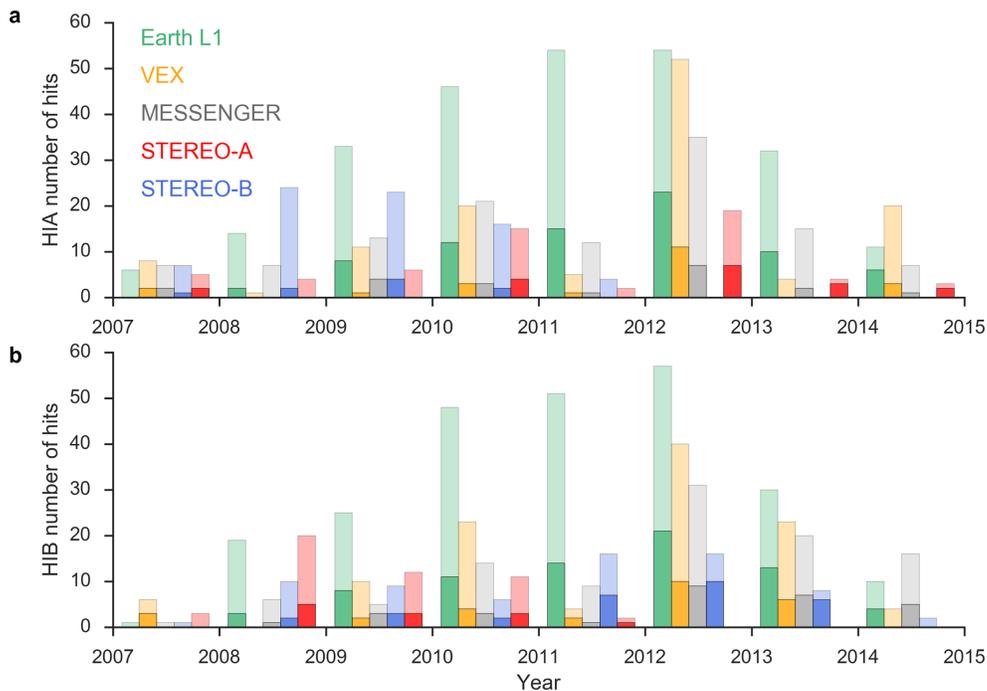

***Figure 3: Comparison of correct predictions and false alarms. (a)*** *The number of predicted CME hits with SSEF30 based on HIGeoCAT for each year (only HIA) is shown as shaded bars as function of time. The solid bars are the number of impacts that are accompanied by an actual ICME in situ detection at the respective planet or spacecraft indicated by the color code. The visible shaded part of the bars indicates the number of false alarms.* ***(b)*** *Same based on HIB observations.*

The solid bars represent the number of impacts that are accompanied by an actual ICME in situ detection at the respective planet or spacecraft indicated by the color code. To this end, a time window of ± 1.5 days was used around the predicted arrival time to search for ICME start times. Table 1 shows the numbers for correctly predicted impacts at each spacecraft and for HIA/HIB separately. The average for MESSENGER and VEX at distances < 1 AU is 22 %, for the other spacecraft close to 1 AU it is 25 % without self-predictions. With self predictions we denote CMEs that were observed by the same STEREO spacecraft with HI and in situ. The accuracy percentage of the self-predictions are significantly higher with 38%, when averaging the numbers from both spacecraft. This underpins the idea that heliospheric imaging might work well also from L1 or Earth orbit [*DeForest and Howard,* 2015*; DeForest et al.,* 2016].

Overall, the prediction accuracy is 26%, so 1 out of 4 predicted impacts actually caused a clear ICME observed in the solar wind. This compares well to *Tucker-Hood et al.* [2015] who found that 1 out of 3, or 20 predicted arrivals out of 60 in total, lead to an ICME observed at Earth/L1, though their time window was larger. If we choose a different time window than 1.5 days, a window of 1.0 days leads for Earth to a percentage of 24%, 0.5 days to 15%, and 2.0 days to 37%, so the true positives range roughly between 1 out of 3 and 1 out of 6, depending which time window is deemed appropriate.

For a time window of 1.5 days, this means that for every correct prediction (solid bars in Figure 3) there are about 3 false alarms (shaded bars). The average for HIB (30 %) is higher than for HIA (23%), which might imply that the "view" from solar east works better compared to solar west, but it is hard to tell if this difference is significant without further analysis. Our study contains a large number of events, i.e., 250 predicted arrivals for Wind could be compared to potential matches within a list of 165 ICMEs, for VEX 121 predictions against 93 ICMEs and for MESSENGER 117 forecasted impacts were compared to 87 ICMEs.

***Table 1:*** *Percentage of correct hits of the ARRCAT predictions, meaning there is an entry in ICMECAT within ± 1.5 days of a predicted arrival in ARRCAT. Fields with * mark "self-predictions". Also shown are mean and standard deviation of arrival time Calculated - Observed, in hours.*

| *Spacecraft* | **Wind** | **STEREO-A** | **STEREO-B** | **VEX** | **MESSENGER** | **Average** |
|---|---|---|---|---|---|---|
| **HIA hits, %** | 30 | 31* | 12 | 17 | 17 | **23** |
| **HIB hits, %** | 31 | 25 | 44* | 25 | 28 | **30** |
| **HIA C-O, hours** | 3.2 ± 16.3 | 1.0 ± 23.4* | 8.0 ± 14.5 | -2.1 ± 11.9 | 3.2 ± 18.1 | **2.4 ± 17.1** |
| **HIB C-O, hours** | 0.8 ± 17.4 | 6.8 ± 16.6 | 7.7 ± 14.7* | -0.7 ± 13.7 | 3.8 ± 12.9 | **2.7 ± 16.0** |

**Figure 4** demonstrates how the percentage of correct hits developed (a) as a function of time and (b) separation from Earth in heliospheric longitude. Here, only the percentages of correct hits at Earth / L1 are shown, for both HIA and HIB. STEREO-B reached the L5 point at 60° separation from Earth in October 2009, and the percentage of correct hits during 2009 was better compared to 2008, where the mean separation was about 30° heliospheric longitude. Surprisingly, the percentage did not decrease with angles larger than L5 but slightly increased as STEREO-A went behind the limb as seen from Earth and finally into conjunction. *Lugaz et al.* [2012] first described the possibility that using HI works for CMEs that propagate behind the limb as seen by an HI spacecraft, and this is consistent with our findings. However, the effect that the percentage of correct hits slightly increases with longitudinal separation needs to be investigated further. The current results imply that L5 is not a particular outstanding point for predicting CME hits and false alarms, but of course still very desirable because of its relative stability.

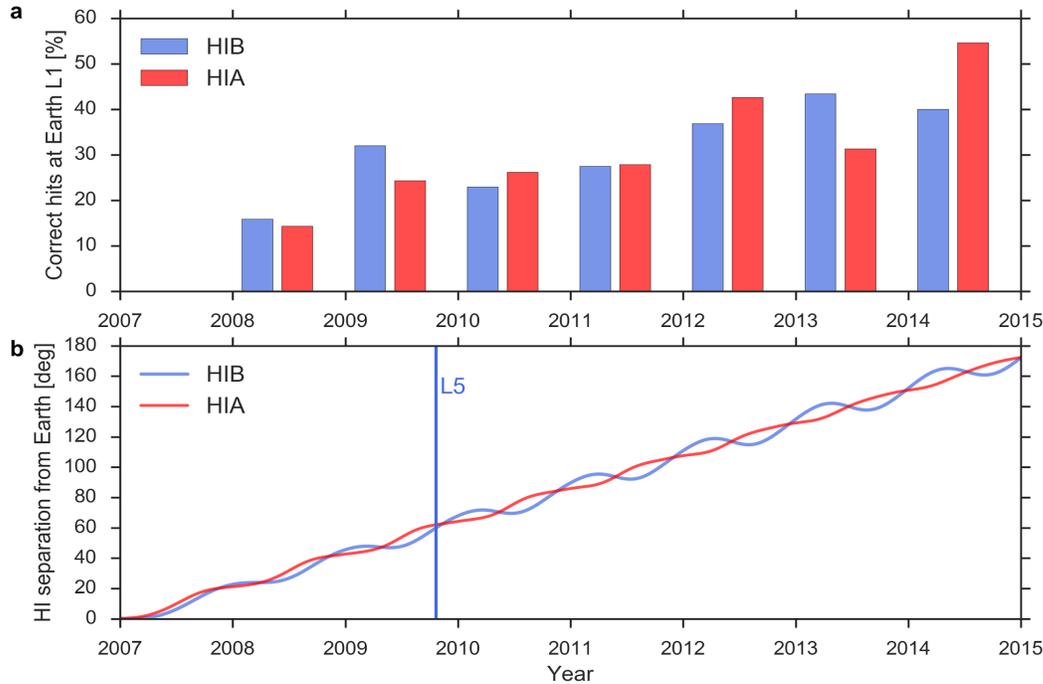

***Figure 4: Correct hit percentage as function of time and longitudinal separation to Earth. (a)*** *Yearly percentage of correct hits as function of time.* ***(b)*** *Separation of STEREO-A/B from Earth in heliospheric (HEEQ) longitude with time. The moment as STEREO-B passed L5 is highlighted as a vertical line.*

### Contingency table results

The common approach in the event-based validation of space weather forecasts is the use of the contingency table [e.g. for CMEs *Zhao and Dryer*, 2014; *Tucker-Hood et al.*, 2015; for flares *Bloomfield* et al., 2012; *Murray et al.*, 2017; and for high-speed streams *Reiss et al.*, 2016]. The performance information contained in the contingency table can be presented in form of verification measures, with the ultimate goal to objectively select the best performing method for operational forecasts. There are some precautions to consider, though. First, the entries in the contingency table need to be clearly defined. Second, different verification measures put different emphases on the table entries and may depend on the event/non-event ratio. Third, a comparison of different methods for CME prediction is only meaningful if the same time range of observations and the same list for in situ observed interplanetary CMEs (ICMEs) is used for CME forecast verification. We would like to emphasize that the catalogs presented here, in particular the ICMECAT containing events at various planets and spacecraft, would be a suitable basis for such investigations with other methods for CME prediction than heliospheric imaging (see "Sources of data and supplementary material" at the end of this study).

With this in mind, in **Table 2** we present the 2 x 2 contingency table for Earth/L1. It contains the numbers of hits (true positives, *TP*) and false alarms (false positives, *FP*), and the values for false rejections (false negatives, *FN*, also called "misses") and correct rejections (true negatives, *TN*). *TPs* are defined by predicted arrivals that are accompanied by an ICME within ± 1.5d of the predicted arrival, *FPs* are predictions that did not result in an ICME (within ± 1.5d of the predicted arrival), and *FNs* are situ ICMEs that were not predicted (defined by no arrival around ± 1.5d of the *icme_start_time*). True negatives are more difficult to assess. *Tucker-Hood et al.* [2015] counted the days where no ICME arrived and none was predicted. Here we take a slightly different approach: we take into account the full length of ± 1.5d as $T_{window}=3d$ because it defined *TP*, *FP* and *FN* before. Thus, we count every $T_{window}$ without an ICME nor a predicted arrival as one correct rejection event, and arrive at the total number of TNs by calculating

$$TN = ( N_{days}/T_{window}) - TP - FP - FN, \qquad (1)$$

with $N_{days}$ the total number of observation days (2697d, HIA; 2736d, HIB). This results in 551/574 (HIA/HIB) intervals of length $T_{window}$ where no CME arrived and none was predicted.

**Table 2:** *Contingency table for Earth / L1 for $T_{window}=3d$. Numbers are given for HIA (HIB).*

| Predicted / Observed | in situ ICME | No in situ ICME |
|---|---|---|
| **ICME arrival** | hit (TP): 76 (74) | false alarm (FP): 174 (167) |
| **no ICME arrival** | false rejection (FN): 98 (97) | correct rejection (TN): 551 (574) |

Based on the contingency table entries several verification measures can be defined [see, e.g., *Woodcock et al.*, 1976; *Jolliffe and Stephenson,* 2006]. Here we focus on the true positive rate (TPR), the false negative rate (FNR), the positive predictive value (PPV), the false alarm ratio (FAR), the threat score (TS), and the bias (BS). The TPR is defined as TP/(TP + FN), the ratio of hits and the total number of observed ICMEs. The FNR is FN/(TP + FN), the ratio of false rejections and the total number of observed ICMEs. The PPV is TP/(TP + FP), the ratio of hits to the total number of predicted ICMEs; and the FAR is FP/(TP +  FP), the ratio of false alarms and the total number of predicted events.

In addition, the threat score (TS) is given by TP/(TP + FP + FN), the ratio of hits to the number of all events. The TS is defined in the range [0,1], where the worst TS is 0, and the best possible TS is 1. The bias (BS) is (TP + FP)/(TP + FN), the ratio of predicted and observed events. The BS indicates if the forecasts are unbiased (BS = 1), tend to under-forecast (BS < 1) or over-forecast (BS > 1) the number of ICMEs. All these measures use parts of the entries in **Table 2**, whereas the Heidke skill score (HSS) and the Hanssen and Kuipers discriminant, also called True skill statistic (TSS), are calculated from the complete contingency table. The TSS and HSS are a measure of the overall forecast performance [for definition, see *Bloomfield et al.,* 2012 or *Jolliffe and Stephenson*, 2006]. We repeat here only the formula for the TSS:

$$TSS = TP/(TP+FN) - FP/(FP+TN). \qquad (2)$$

The TSS is defined in the range [-1,1], where the worst TSS is 0 and the best possible TSS is 1 or -1 (inverse classification). In contrast, the HSS is defined in the range [-∞, 1], where the worst HSS is smaller than 0 (HSS = 0, indicates no forecast skill) and the best HSS is 1.

**Table 3:** *Skill scores for Earth/L1 for predictions with the SSEF30 technique with heliospheric imagers and verified with the Wind interplanetary CME (ICME) list. Valid for a time window of ± 1.5d. Total number of CME events: 697 (HIA), 653 (HIB); total predicted arrivals: 250 (HIA), 241 (HIB); total observed ICMEs at Earth/L1: 165.*

| Skill score | short form | results HIA | results HIB |
|---|---|---|---|
| True positive rate | TPR | 0.44 | 0.43 |
| False negative rate | FNR | 0.56 | 0.57 |
| Positive predictive value | PPV | 0.30 | 0.31 |
| False alarm ratio | FAR | 0.70 | 0.69 |
| Threat score | TS | 0.22 | 0.22 |
| Bias | BS | 1.44 | 1.41 |
| Heidke skill score | HSS | 0.17 | 0.18 |
| True skill statistic | TSS | 0.20 | 0.21 |

In **Table 3**, we present the skill scores computed for Earth/L1. We find that all verification measures computed are similar for HIA and HIB, respectively. For HIA (HIB) we find that the TPR is 0.44 (0.43), the FNR is 0.56 (0.57), the PPV is 0.30 (0.31) and the FAR is 0.70 (0.69). This means that 44% (43%) of ICMEs are correctly predicted with the SSEF30 technique, while only 30% (31%) of all ICMEs forecasted are observed. The computed BS of 1.44 (1.41) also confirms that the total number of ICMEs is clearly over-forecasted. Focusing on the TS we find that 22% (22%) of ICME forecasts amongst all ICME events (either forecasted or observed) were correct. The high numbers of false alarms and misses are also reflected in the computed HSS and TSS, where the HSS is 0.17 (0.18) and the TSS is 0.20 (0.21).

Regarding the dependence on the length of the time windows, it is very interesting to note that the TSS is independent of the time window between ± 1.0d and ± 2.0d, resulting in very little variation of TSS within 0.19 to 0.22. Outside of this range of time windows, the TSS declines.

Considering the differences in the setup used (HI method, ICME list, and time window) our results are comparable with *Tucker Hood et al.* [2015] who report a TPR of 0.31, a FAR of 0.64 and a HSS of 0.27.

The TSS for STEREO-A self predictions (i.e. HIA verified by STEREO-A in situ observations) is only 0.05, because of an extremely high FNR of 0.90. This means most CMEs that are detected by the in situ instruments are missed by its heliospheric imager. For STEREO-B most skill scores and in particular the FNR are comparable to STEREO-A, though TSS is here slightly better with a TSS of 0.15.

### Arrival times, speeds and magnetic fields

**Figure 5** demonstrates the arrival time differences for each spacecraft and HIA/HIB separately (left and right column, respectively). We show the calculated (C) minus observed (O) differences in arrival times in hours. Positive values stand for cases where the CME arrived earlier than the forecast implied, and negative values signify late CME arrivals. Again we use a time window of ± 1.5 days around all predicted arrival times in ARRCAT. The choice of 1.5 days is somewhat arbitrary, so we also quote results for ±1.0 and ±2.0 days. The ICME start time that is closest to the predicted arrival is taken for the C-O calculation, and this time difference must be inside the time window for which the C-O is calculated. All other ICMEs inside and outside this time window are ignored. These "double hits", where a predicted CME impact could be related to two ICME start times, are quite seldom and happen, e.g., at Earth only for 12% (9%) of arrivals predicted using HIA (HIB).

The total number of comparisons for HIA was 171, for HIB 194. The average C-O is 2.4 ± 17.1 hours for HIA, and 2.7 ± 16.0h for HIB, so no significant difference. For both spacecraft the average is 2.6 ± 16.6h. For Earth/L1 alone, this result is 3.2 ± 16.3h (HIA) and 0.8 ± 17.4h (HIB). With a time window of 1.0 days, this result is 2.3 ± 12.5h (HIA), 0.0 ± 12.1h (HIB), and for 2.0 days its 5.3 ± 22.5h (HIA), and 1.3 ± 22.9h (HIB). This means that an arrival time difference depends on the time window and thus, a time window should always be given when discussing arrival time differences.

In comparison, *Möstl et al.* [2014] find an average C-O = -1.4 ± 11.1 (their Table 3) for the SSEF30 technique, with all in situ spacecraft at 1 AU. The difference is explained by the selection of events by these authors that quite clearly matched between HI and in situ, whereas here there is no such selection. Most values in Table 2 are positive, so the CME arrived slightly earlier than predicted.

As an experiment, we have excluded events that propagate too far away from the solar equatorial plane in latitude so the only CMEs with a central position angle (PA) between ± 20 degree around a PA of 90° (HIA) or 270° (HIB) are included as impacts. This works slightly better concerning the hit and miss percentages with 27% (HIA, a plus of 4%) and 35 (HIB, plus 5 %). The values for the arrival time differences are overall here 2.8 ± 15.5h (HIA) and 2.6 ± 15.2h (HIB), with no significant change in the averages and improvements by 1.6 (HIA) and 0.8 (HIB) hours in the standard deviations.

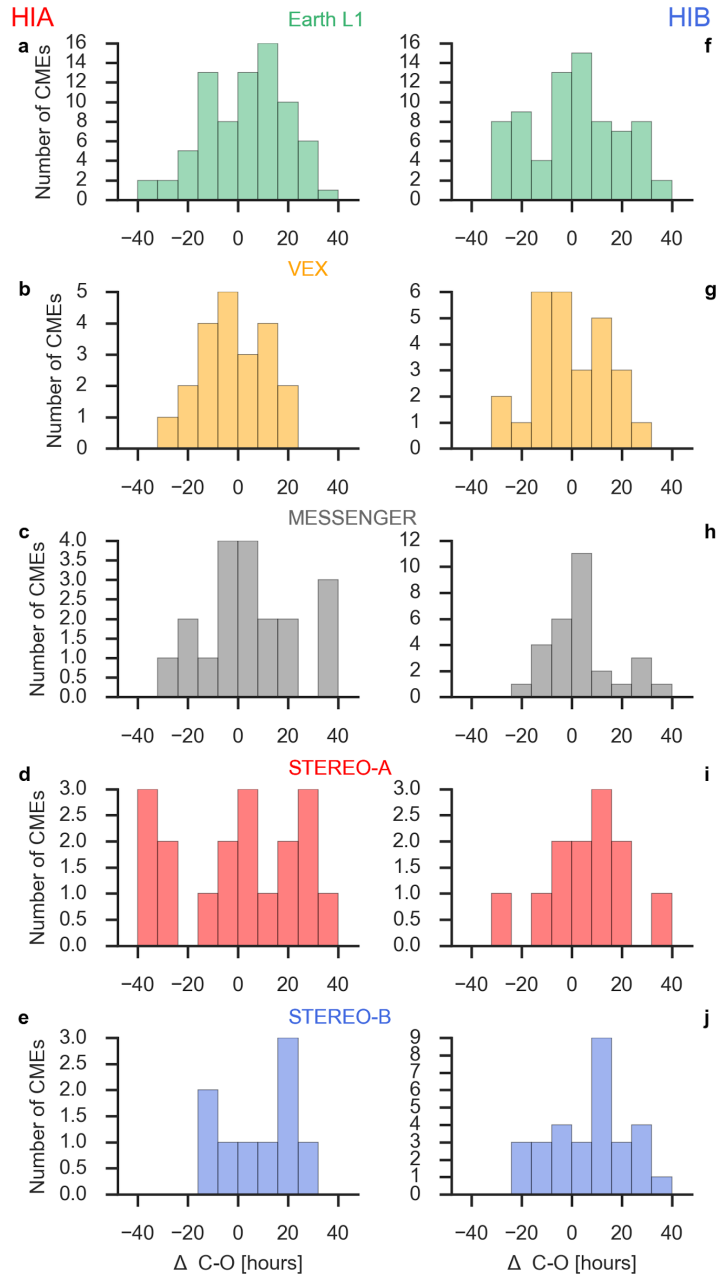

**Figure 5: Histograms of calculated (C) minus observed (O) differences in arrival times, in hours for each spacecraft.** *From top to bottom: Wind, VEX, MESSENGER, STEREO-A, STEREO-B Left column (a-e) shows observations based on HIA, right based on HIB (f-j).*

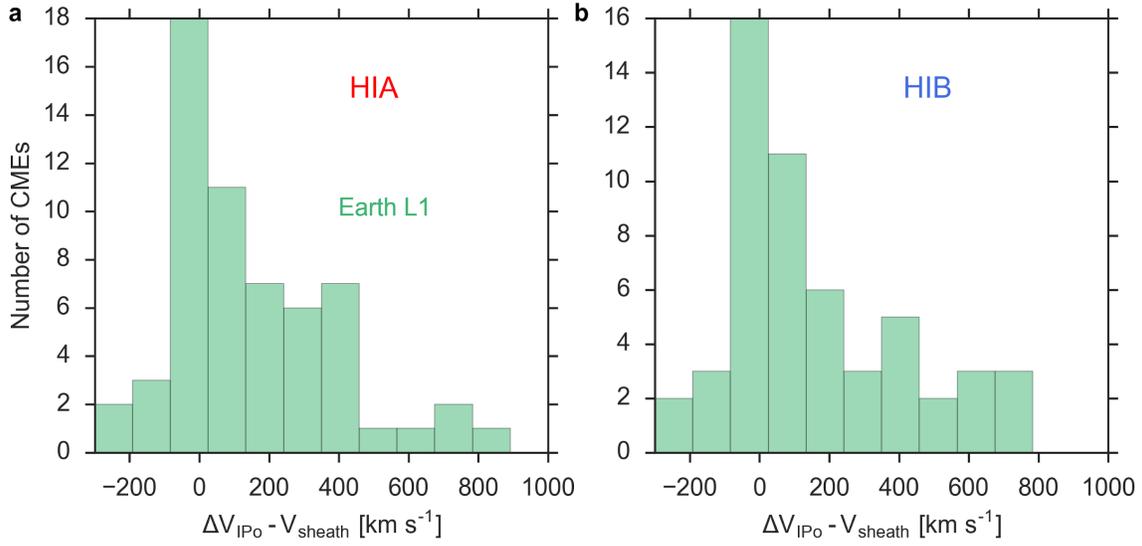

***Figure 6: Difference of HI predicted speeds to ICME plasma speeds. (a)*** *Calculated minus observed CME speed at Earth, for HIA. The HI speed is corrected for the circular SSEF30 front. The in situ speed is the proton speed in the ICME sheath region.* ***(b)*** *Same for HIB.*

**Figure 6** demonstrates the calculated minus observed CME speed at Earth. The in situ speed for comparison is the proton speed in the ICME sheath region. It is seen that many events cluster around a speed difference of 0, but there is a positive tail to the distribution, which is a consequence of the constant speed assumption of SSEF30, overestimating the arrival speed [*Möstl et al.,* 2014]. The HI speed is corrected for the circular SSEF30 front [*Möstl and Davies,* 2013]. The in situ speed for comparison is the proton speed in the ICME sheath region. The differences are for HIA at Earth L1: 191 ± 341 km s⁻¹, and HIB at Earth L1: 245 ± 446 km s⁻¹. These results are very similar to *Möstl et al.* [2014] who quote 252 ± 302 km s⁻¹. Because we do not have plasma speeds from VEX and MESSENGER we cannot do this comparison for < 1 AU.

It is also of great interest to find constraints on the magnitude of in situ ICME magnetic fields from the HI data. **Figure 7** visualizes two relationships between parameters derived from HI as part of the arrival catalog and the in situ mean magnetic field in the magnetic obstacle.

**Figure 7a** shows the predicted speed at the target and the mean magnetic field in the magnetic obstacle, between *mo_start_time* and *mo_end_time*. This is plotted for the hits (true positives) in the dataset. Most events cluster around slow to intermediate speeds (< 1000 km s⁻¹) and low magnetic fields (< 20 nT). Only at MESSENGER, because of its closer radial distance to the Sun, high average magnetic fields of up to about 90 nT are seen, at all other spacecraft they have been quite low during this solar cycle. For the events classified as hits, the averages ± standard deviations in situ magnetic fields in the magnetic obstacles are: for Earth/L1 10.5 ± 5.0 nT, for VEX at Venus 19.8 ± 9.1 nT, and for MESSENGER 43.3 ± 17.9 nT. For a few events where the HI-in situ connection was quite clear and only in situ events near 1 AU were used, *Möstl et al.* [2014] derived a linear relationship for a similar plot. In Figure 7b, the predicted travel time, which is defined as *target_arrival - sse_launch,* is plotted against the mean magnetic field. This highlights that only short travel times are accompanied by strong magnetic fields. This relationship does depend on radial distance, as the events from MESSENGER again dominate the left side of the plot.

These plots and values may give some rough indication of the magnitude of the magnetic field that is to be expected when looking at the predicted CME travel time, the predicted impact speed and its heliospheric location.

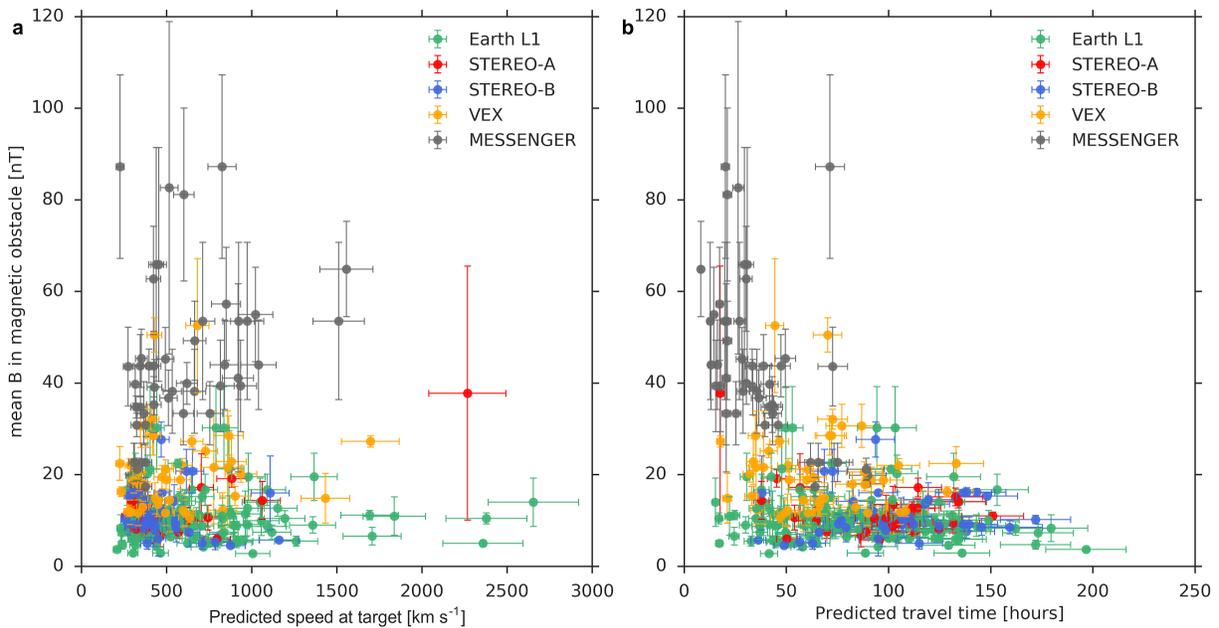

*Figure 7: Relationship between predicted speed, travel time and the mean magnetic field in the magnetic obstacle (MO), for all events classified as hits (true positives). (a) The predicted target speed is shown with a typical error bar of ± 10% [Möstl et al., 2014] against the mean MO field, with the error bar given by the standard deviation of the MO field. (b) The predicted travel time, also with a ± 10% error, against the mean MO field. The color code for each spacecraft is given in the plot legend.*

## 4 Conclusions

We have compared the predictions of the impact and the arrival time from the SSEF30 technique using STEREO/HI observations for more than 1300 CME events with more than 600 ICMEs that were reported in the existing literature and online data sources and merged by us into a singular catalog. Thereby we have derived some fundamental results concerning heliospheric imager-based geometrical modeling with the SSEF30 technique. This is of major importance for a space weather mission carrying a heliospheric imager at the L5 or L1 point *[Lavraud et al.*, 2016, *DeForest et al.* 2016], and all other missions that carry heliospheric imagers (*STEREO, Solar Orbiter, Solar Probe Plus*). In summary, our results derived from science data and those by *Tucker-Hood et al.* [2015] for HI real time data show that HI modeling is currently at the same level for predicting CME arrivals as are other analytical and numerical techniques.

**Contingency table results:** We find that 1 out of 4 predicted impacts with the SSEF30 technique results in a clear ICME that includes a magnetic obstacle, for a time window of ± 1.5 days around the predicted arrival time. Including a constraint on the propagation latitude of the CME in HI, so that only CMEs which centrally propagate near the solar equatorial plane are selected, improves the hit percentages to about 33 %, so 2 false alarms for 1 correct prediction.

We find for time-windows from 1.0 to 2.0 days a stable true skill statistic (TSS) of 0.19 to 0.22, similar for HIA and HIB. What does this mean in comparison to other studies? We need to caution that we have used science data to discuss hindsight predictions of CMEs with heliospheric imagers. *Tucker-Hood et al.* [2015], analysing the real time performance of HI with beacon data, find a slightly higher TSS of 0.26 (calculated from their contingency table). The higher TSS than in our study is likely explained by smaller statistics and a longer time window as in our study. For several techniques to calculate the shock arrivals, *Zhao and Dryer* [2014] find TSS between 0.15 and 0.25, similar to our results.

Overall, the TSS of about 0.20 might be considered rather low concerning the accuracy of a regular space weather forecast. However, we recognize several aspects that are expected to improve the prediction accuracy. First, the HI SSEF30 technique does not adequately reflect basic CME physics (see below). Secondly, our analysis relied on existing ICME lists, focusing on clear ICMEs that feature magnetic flux ropes. We did not include ICME flank encounters that often feature unclear ICME signatures [e.g., *Richardson and Cane,* 2010] and sometimes "driverless shocks" [e.g., *Gopalswamy et al.,* 2010; *Janvier et al.,* 2014], i.e., the whole bulk of the CME is missed. Even such cases are interesting for space weather as CME sheaths alone can drive major storms [*Huttunen et al.,* 2002; 2004].

In future work, the forecasted CME arrivals might be checked against the in situ data directly, not just the ICME lists, which would eliminate the identification of ICMEs which is always a subjective matter. Another clear improvement is to take into account event specific aspects. A major assumption that needs to be eliminated is the constant 60° full width of the CME [based on *Yashiro et al.,* 2004], and determine the width for each event separately. This emphasises also the importance of multi-spacecraft missions to allow CME reconstructions from multiple vantage points to obtain realistic parameter that are not subject to projection effects [e.g., *Thernisien, Vourlidas and Howard*, 2009].

**Arrival times:** Forecasting with HI also works well for locations at < 1 AU, as we find that the predicted arrival times match with similar accuracy compared for spacecraft positioned around 1 AU. This result might be expected for Mars too, though we could not yet test this as MAVEN entered its orbit around Mars in late 2014 just when STEREO went into conjunction. The mean, signed arrival time differences of around ± 17 hours are very close to those reported by a comparison of the heliospheric MHD simulation Enlil and the drag-based model (DBM) forecasts at Earth [*Vršnak et al., 2014*]. Our sample for this comparison consists of 315 CMEs, 143 observed by HIA and 172 by HIB. *Mays et al.* [2015] report a mean absolute difference of 12.3 hours for predicted shock arrivals with Enlil at Earth; *Zhao and Dryer* [2014] quote 10 hours. For our study and looking only at the Earth/L1 arrivals, this mean absolute difference is only slightly larger: 14.0 (HIA) and 14.4 hours (HIB), based on 76 (HIA) and 74 (HIB) comparisons to in situ data. However, it must be noted that such comparisons depend on the choice of the time window for successful hits. In summary, with the current SSEF30 technique, using an HI instrument does not enhance the CME prediction accuracy, but its results are perfectly comparable with those from other methods that use analytical and numerical modeling.

To improve the arrival time predictions, in future work, the SSEF30 modeling results on all 1337 CMEs should be updated with ElEvoHI [*Rollett et al., 2016*], as the constant speed of SSEF30 assumption runs into troubles in particular predicting the arrival speeds. A robust automatization of ElEvoHI is needed to do this. ElEvoHI eliminates both assumptions of a circular front and constant speed, and instead uses an elliptical shape [*Janvier et al.*, 2014; *Möstl et al.*, 2015; *Rollett et al.*, 2016] and a decelerating speed profile [*Vršnak et al.*, 2013]. The ellipse shape may even evolve with time. We expect that applying ElEvoHI to all CME tracks in the HIGeoCat will show improvements concerning the arrival time and arrival speed. Future work should also focus on adding magnetic fields to predictions with heliospheric imagers, by bringing together a model such as FRi3D [*Isavnin, 2016*] or FIDO [*Kay et al., 2017*] with ElEvoHI to make predictions of CME geomagnetic effects possible [*Tobiska et al.,* 2013; *Kubicka et al.*, 2016].

**Implications for future missions:** We have demonstrated that L5 could be a good location for the prediction of earth-directed CMEs with HI. But somewhat surprisingly, if the HI observing spacecraft is positioned further away from the Earth in heliospheric longitude, the hit/miss predictions do not get worse. Note, however, that this was already seen by *Lugaz et al.* [2012] and *Möstl et al.* [2014]. We can now base this conclusion on a larger event sample, but there are other factors such as a changing solar activity from minimum to maximum that may considerably influence this result, and more specific work should be dedicated to understanding this effect. In any case, this has an implication for current operations with STEREO-Ahead: real time predictions with HI should be done even when the spacecraft is still behind the east limb of the Sun as seen from Earth. This means we do not have to wait to use HI until it passes L5 in July 2020 to make good predictions using the HI instrument.

It has also been proposed that L1 could be a good location for a heliospheric imager [*DeForest and Howard*, 2015]. While we find that "self-predictions" of CMEs observed with HIA or HIB and later detected with the same spacecraft by the in situ instruments measuring magnetic fields and plasma parameters show slightly better percentages of correct hits than those to other spacecraft, the rate of false rejections is much higher than for an "away" view-point, meaning most ICMEs that impacted STEREO-A or B were not detected by its own heliospheric imager.

**Acknowledgments, Samples, and Data**

The presented work has received funding from the European Union Seventh Framework Programme (FP7/ 2007-2013) under grant agreement No. 606692 [HELCATS]. This study was supported by the Austrian Science Fund (FWF): [P26174-N27]. L.P. acknowledges support from the Natural Sciences and Engineering Research Council (NSERC) of Canada. V. K. acknowledges the support of the Czech Science Foundation grant 17-06818Y. This research has made use of SunPy, an open-source and free community-developed solar data analysis package written in Python [*The Sunpy Community et al.*, 2015]. We would like to thank the editor and two reviewers for helpful suggestions to improve the manuscript.

All sources of data that were used in producing the results presented in this study are quoted below.

**Sources of data and supplementary material**

Catalogs:

HIGeoCat:  https://www.helcats-fp7.eu/catalogues/wp3_cat.html

ARRCAT: doi:10.6084/m9.figshare.4588324

https://doi.org/10.6084/m9.figshare.4588324.v1

ICMECAT: doi:10.6084/m9.figshare.4588315

https://doi.org/10.6084/m9.figshare.4588315.v1

ICME lists:

Wind ICME list: T. Nieves-Chinchilla et al. https://wind.nasa.gov/ICMEindex.php

STEREO ICME list: Lan Jian, http://www-ssc.igpp.ucla.edu/forms/stereo/stereo_level_3.html

updated by our study.

VEX: *Good and Forsyth* [2016], updated by our study.

MESSENGER: *Winslow et al.* [2015], *Good and Forsyth* [2016], updated by our study.

In situ data:

Wind: https://cdaweb.sci.gsfc.nasa.gov

STEREO: http://aten.igpp.ucla.edu/forms/stereo/level2_plasma_and_magnetic_field.html

MESSENGER: http://ppi.pds.nasa.gov/search/?sc=Messenger&t=Mercury&i=MAG

VEX: data obtained from magnetometer PI T.L. Zhang Tielong.Zhang@oeaw.ac.at

Animation of Figure 2:

doi.org/10.6084/m9.figshare.4602253

https://www.youtube.com/watch?v=Jr4XRzGCaaQ&